\documentclass[prb,twocolumn,showpacs,preprintnumbers,
     floatfix,superscriptaddress]{revtex4}
\usepackage{graphicx}
\usepackage{dcolumn}
\usepackage{bm}
\usepackage{amsmath}
\usepackage{amsfonts}
\usepackage{amssymb}

\begin{document}

\title{Nonlinear elastic and electronic properties of Mo$_6$S$_3$I$_6$
nanowires}
\author{I. Vilfan}
\author{D. Mihailovic}
\affiliation{J. Stefan Institute, Jamova 39, SI-1000 Ljubljana, Slovenia}
\date{\today}

\begin{abstract}
The properties of  $\mathrm{Mo_6S_3I_6}$ nanowires were investigated with ab
initio calculations based on the density-functional theory. The molecules 
build weakly coupled one-dimensional chains with three sulfur atoms in the
bridging planes between the Mo octahedra, each dressed with six iodines.  
Upon uniaxial strain along the wires, each bridging plane shows two energy 
minima, one in the ground state with the calculated Young modulus  $Y=82$
GPa, and one in the stretched state with  $Y=94$ GPa. Both values are at least
four times smaller than the experimental values and the origin of the discrepancy
remains a puzzle. The ideal tensile strength is about 8.4 GPa, 
the chains break in the Mo-Mo bonds within the octahedra and not in the S
bridges. The charge-carrier conductivity is strongly anisotropic and 
the $\mathrm{Mo_{6}S_{3}I_{6}}$ nanowires behave as quasi-one-dimensional 
conductors in the whole range of investigated strains. The conductivity  is extremely 
sensitive to strain, making this material very suitable for stain gauges.
Very clean nanowires with good contacts may be expected to behave as
ballistic quantum wires over lengths of several $\mu $m. On the other hand, with
high-impedance contacts they are good candidates for the observation of
Luttinger liquid behaviour. The pronounced 1D nature
of the $\mathrm{Mo_{6}S_{3}I_{6}}$ nanowires makes them a uniquely versatile 
and user-friendly system for the investigation of 1D physics.
\end{abstract}

\maketitle



\section{Introduction}

Inorganic nanowires and nanotubes are rapidly gaining in importance as
materials 
of great scientific and technological interest because they exhibit a
substantially enhanced level of functionality which can open novel
opportunities for devices and applications. The main advantage of inorganic
nanowires over other related materials is the possibility of controlling the
relevant physical properties by selective engineering of their geometry
and/or composition. Molybdenum-based inorganic nanowires and nanotubes in
particular are emerging as materials with very promising physical properties
because they are easier to functionalize and to synthesize in clean, single
stoichiometry form than other one-dimensional materials. The synthesis, some
physical properties and applications of several molybdenum-based nanowires
have already been reported.\cite{PCS80,Tarascon_84,BMPGS88,VL99, RRC02}
Recently, significant research effort has been focused on nanowires composed
of Mo, chalcogens (S) and halogens (I) in the form $\mathrm{%
Mo_{6}S_{9-x}I_{x}}$ (MoSI$_{x}$).\cite{MKPG05,YOBT06} which are best
described as molybdenum chalcogenide-halide polymers. They have a backbone
skeleton composed of Mo$_{6}$ octahedra, each dressed with six anions (S or
I), bound together into one-dimensional chains by three bridging anions (either S or
I). A specific feature of these materials is the growth of identical chains
in the form of bundles. The materials have strong anisotropy, large Young
moduli along the wires, very small shear moduli, and controllable electronic
properties. In contrast to related nanowires such as Li$_{2}$Mo$_{6}$Se$_{6}$
which decompose rapidly in air, the MoSIx nanowires are entirely air-stable.
Yang et al.\cite{YOBT06} have recently investigated the $\mathrm{%
Mo_{12}S_{9}I_{9}}$ nanowires theoretically and have pointed out the
bistable nature of the sulfur bridges, which allow the wires to stretch in
an accordion-like fashion at virtually no energy cost. They also pointed
that strain can have a significant effect on the electronic band structure,
and, interestingly, the appearance of a net spin polarisation within the
nanowires in the unstrained position.

\begin{figure}[tbp]
\includegraphics[width=0.6\columnwidth]{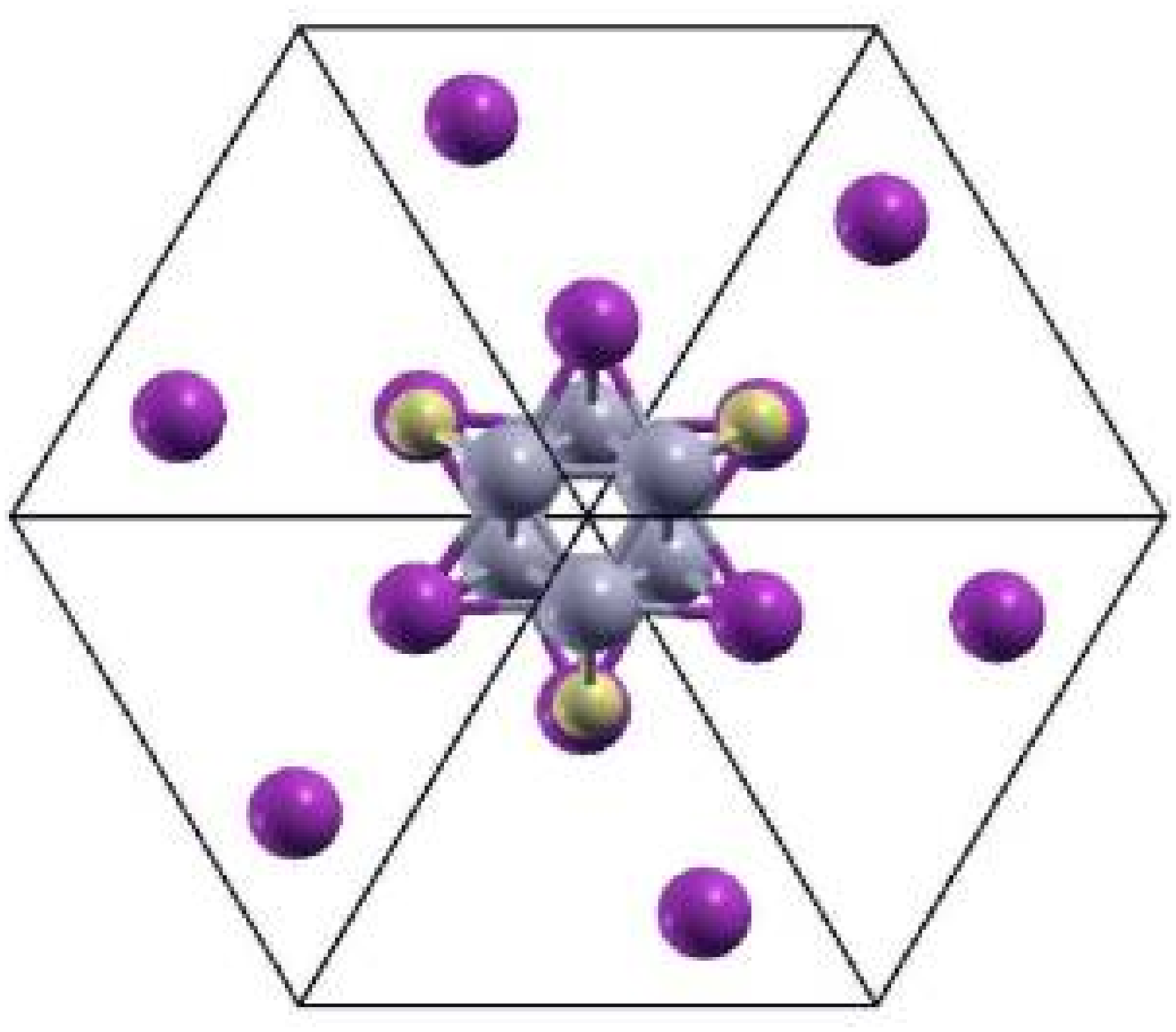}\hskip1cm %
\includegraphics[width=0.6\columnwidth]{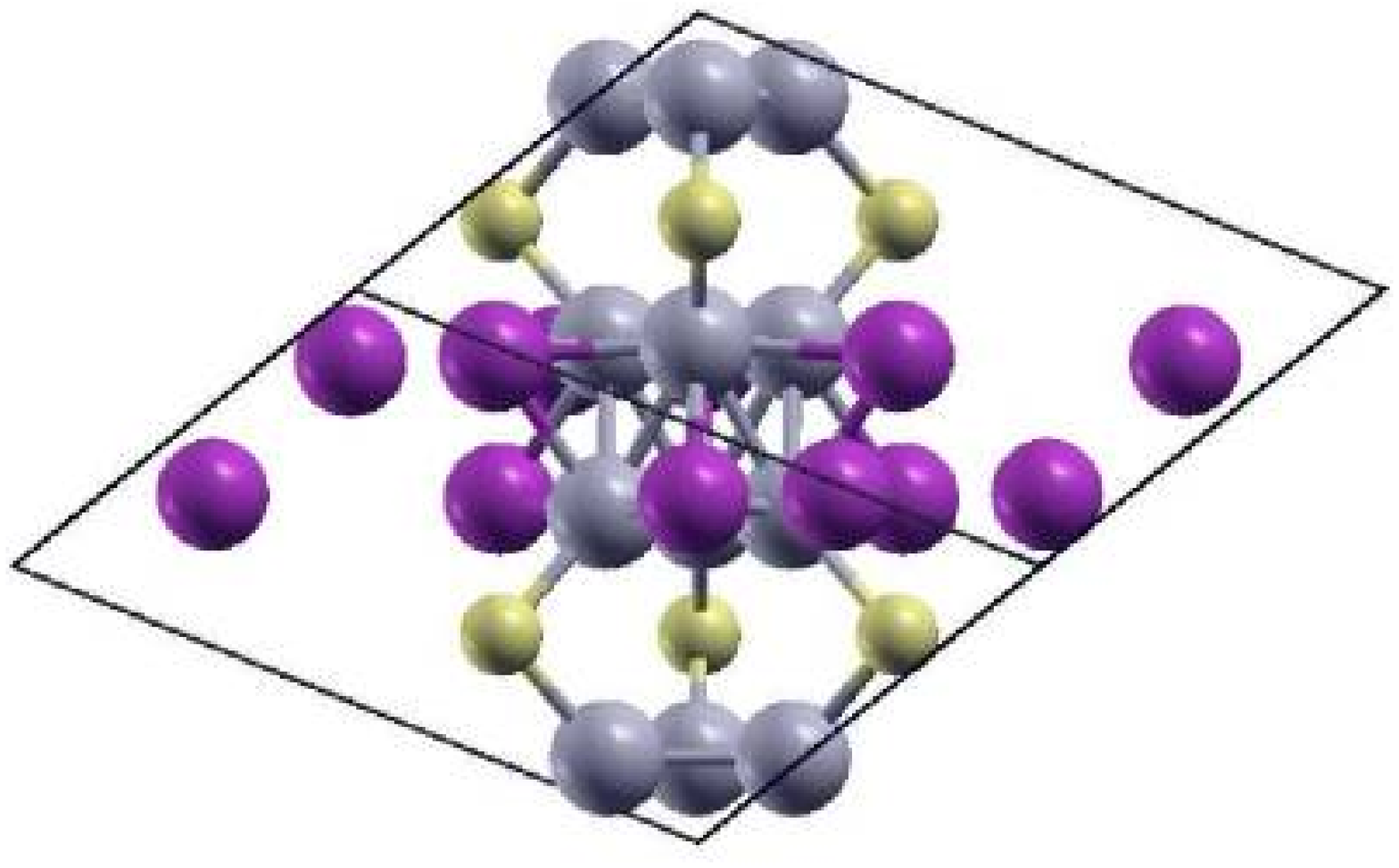}
\caption{(Color online) Top (upper panel) and side (bottom panel) view on
the primitive cell of $\mathrm{Mo_6S_3I_6 } $ nanowires ordered according
to the spacegroup R$\overline{3}$c. Mo octahedra (grey) are dressed with 6 I
atoms (violet - dark grey) and separated by 3 bridging S atoms (yellow -
light grey). The X-ray scattering data and the DFT calculations of the R%
$\overline{3}$ isomer indicate a $\sim6^\circ$ rotation of the chains around
the hexagonal $c$ axis and a certain degree of occupational (either S or I)
disorder. The distant I atoms belong to the neighbouring chains.}
\label{structure167}
\end{figure}

In this paper we report on the theoretical investigations of
extraordinary elastic and electronic properties of the most common
reproducibly synthesized MoSI$_x$ nanowire material, namely $\mathrm{%
Mo_{6}S_{3}I_{6}}$. 
Meden et al.\cite{MKPG05} reported the structure where,
after purification, the X-ray diffraction experiments 
indicated an iodine occupancy of the bridging atoms and only a small 
degree ($\sim 10$ \%) of disorder in the anion occupancies (either S or I).
More recently, Nicolosi et al.\cite{STEM} reported on the electron microscopy
of same isomer with the same skeletal structure but with the bridging
planes occupied by the sulfur atoms and all the iodines attached to the Mo
octahedra. As we will comment later, both isomers have almost the same energy
therefore it is possible that both are synthesized at similar rates. 
The question arises, how to identify them, what are the similarities and what
are the differences between the two isomers. We shall concentrate in this paper
on the isomer with three S atoms in the bridging planes, 
Fig.~\ref{structure167}. 
From the tribology
experiments we know that the interaction between individual nanowires is
weak,\cite{JDM05} therefore we focus our analysis on the longitudinal
properties of individual nanowires, ordered in a hexagonal array according
to the spacegroup R$\overline{3}$c. 
This space group assumes the inversion symmetry which makes all the sulfur
bridges equivalent. As a consequence we can scan the whole range of strains.
Using the density-functional theory (DFT) we investigate the effect of
strain on the elastic and electronic properties of these nanowires.


\section{Computational Details}

The structure and properties of Mo-S-I nanowires were analysed using the
WIEN2k code.\cite{BSMKL01} The simulations were performed on a mixed basis
set of augmented plane waves plus local orbitals (APW+lo)\cite{SNS00} for
low orbital momenta ($\ell \le 2$) and linearized augmented plane waves
(LAPW) for all the higher orbital momenta. The exchange and correlation
potential was treated in the 
Perdew, Burke and Ernzerhof generalized-gradient approximation.\cite{PBE96}
The muffin-tin radii were set to $R_{\mathrm{Mo}}^{\mathrm{MT}} = 1.06$ {%
\AA\ } for Mo, 
$R_{\mathrm{S}}^{\mathrm{MT}} = 1.27$ {\AA } for S, and 
$R_{\mathrm{I}}^{\mathrm{MT}} = 1.38$ {\AA } for I, 
the kinetic energy cutoff was $E_{\mathrm{max}}^{\mathrm{wf}} = 12.3$ Ry and
the plane-wave expansion cutoff was 
$E_{\mathrm{max}}^{\mathrm{pw}} = 196 $ Ry. 

First the hexagonal lattice constants $a=b$, $c$ and the atomic coordinates
were optimized. In this stage the energy was calculated on a tetrahedral
mesh with 44 $k$-points in the irreducible part of the Brillouin zone. 
The same density of $k$-points was used for the analysis of elastic
properties. Later, when calculating the electronic density of states, 189 $k$%
-points were used. 
Since the interactions between neighbouring chains are weak, we kept the
lattice constants $a$ and $b$ fixed, varied $c$ and optimised the atomic
coordinates when calculating the properties of strained crystals. 
We thus obtain the elastic constants $c_{11}$ which are close to the Young
moduli $Y$ because of weak interchain coupling.


\section{Results}

\begin{figure}[tbp]
\includegraphics[width=0.8\columnwidth]{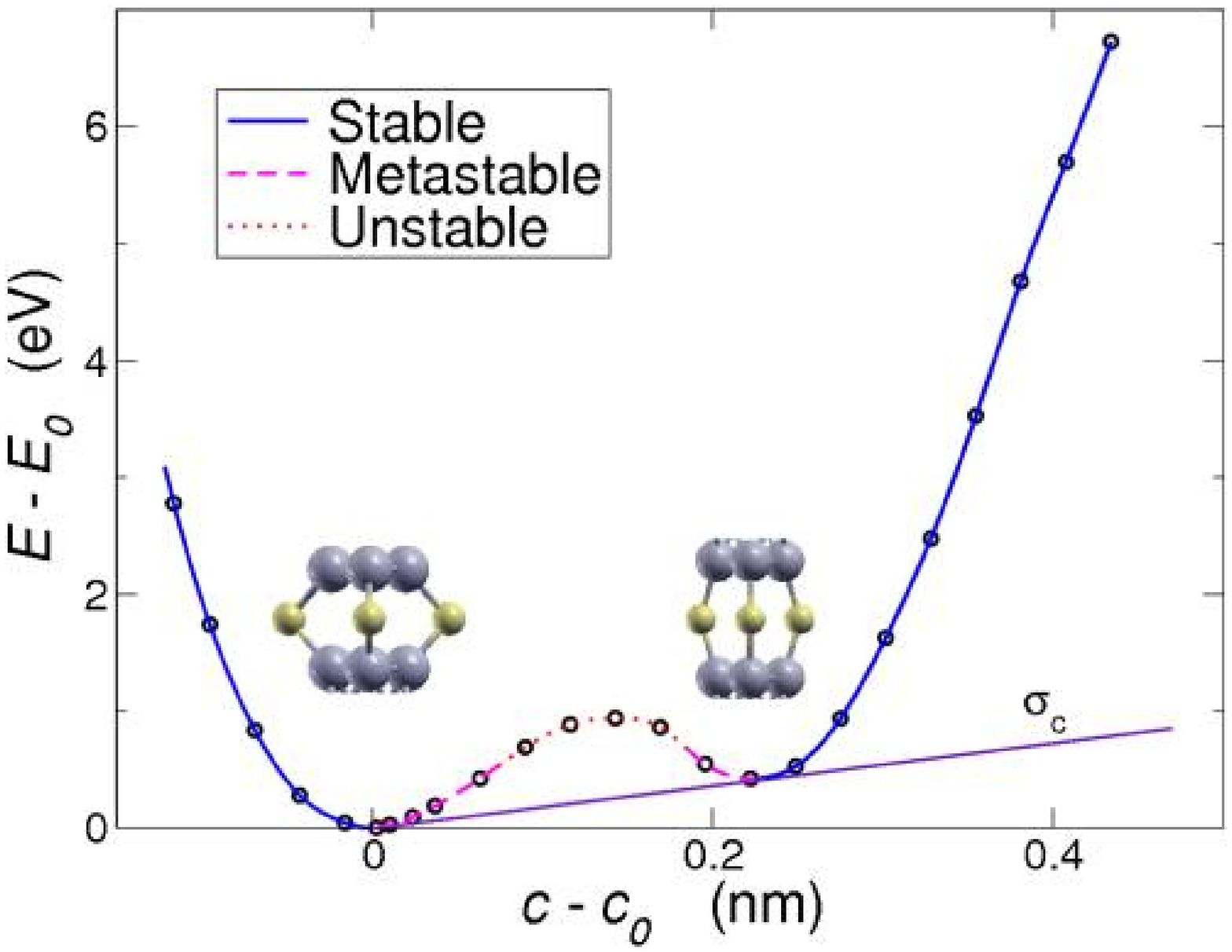}
\begin{center}
(a)
\end{center}
\includegraphics[width=0.8\columnwidth]{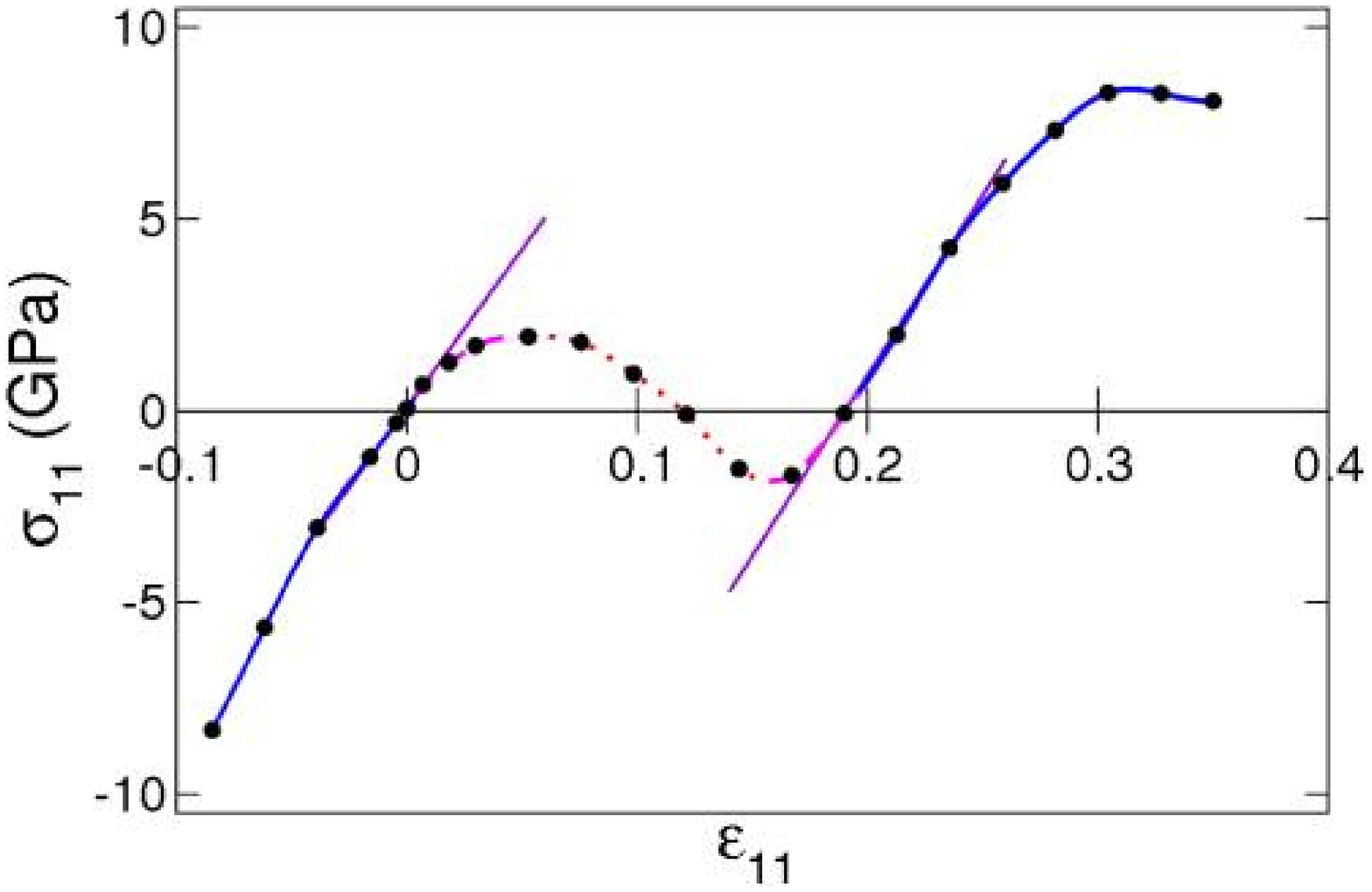}%
\newline
(b)
\caption{(Color online) (a) Strain dependence of the binding energy per
primitive cell (two $\mathrm{Mo_6S_3I_6}$ molecules). The two minima
correspond to short and long Mo-S-Mo bridge configurations, respectively. In
the unstable region the wire has short and long Mo-S-Mo bridges
simultaneously. The inserts show the Mo-S-Mo bridges in the two energy
minima. The straight line indicates the critical stress, i.e., the
equilibrium transition between the two minima. (b) Stress-strain graph of $%
\mathrm{Mo_6S_3I_6}$ nanowires. The slopes of straight lines indicate
the elastic constants of the two stable configurations. The maximum at $%
\protect\epsilon_{11} = 0.315$ is the ideal tensile strength of $\mathrm{%
Mo_6S_3I_6}$ nanowires with all the Mo-S-Mo bridges in the long
configuration.}
\label{stress167}
\end{figure}

The DFT simulations of Mo-S-I arranged as shown in Fig.~\ref{structure167}
give the equilibrium hexagonal lattice constants $c=11.56$ {\AA\ } and $%
a=16.54$ {\AA } 
which agree to within 3\% with the experimental values $c=11.95$ {\AA } and $%
a=16.39$ {\AA } although we assumed a different spacegroup (the simulated
group was R$\overline{3}$c, while the X-ray diffraction suggests P$6_{3}$ or
lower).\cite{MKPG05} In this configuration the Mo-S-Mo angle across the
bridging plane is $90.8^{\circ }$. Upon stretching the nanowires we find
another energy minimum at $c=13.76$ {\AA } with the Mo-S-Mo angles $%
145^{\circ }$, Fig.~\ref{stress167}(a). Yang et al.\cite{YOBT06} ascribed
the two equilibrium configurations to different s-p hybridizations of
sulfur. However, inspection of the atomically resolved electron density of
states, discussed later, shows very little s-p hybridization on sulfur. On
the other hand, in the second energy minimum, the separation between the S
atoms in the same bridging plane, $d_{S-S}=3.79$ {\AA }, is close to twice
the van der Waals radius ($r_{vdW}=1.85$ {\AA }) or to the separation between
S in neighbouring layers in layered MoS$_{2}$, $d_{vdW}=3.49$ {\AA }. 
Therefore we suggest the possibility that the van der Waals interaction 
between S on the same bridging plane stabilizes the second energy minimum.

The deeper minimum (unstrained wire) is subject to a potential barrier $%
\Delta E = 0.47$ eV and the higher one by $\Delta E = 0.26$ eV, both per
formula unit. These barriers are too high to be thermally excited, the only
way to move from one to the other minimum, is by applying an external stress.


\subsection{Elastic properties}

Like other MoSI$_x$ nanowires, also the $\mathrm{Mo_{6}S_{3}I_{6}}$
nanowires have large interchain separations with weak, van der Waals
coupling between the chains. As a consequence, the nanowires are very
anisotropic in their properties, including elasticity. They have large
elastic constants in the direction of the wires, $c_{11}$, and small shear
and Young moduli perpendicular to the wires. Here we shall concentrate on
the most interesting elastic constant, $c_{11}\approx Y$. 
\begin{figure}[tbp]
\includegraphics[width=0.8\columnwidth]{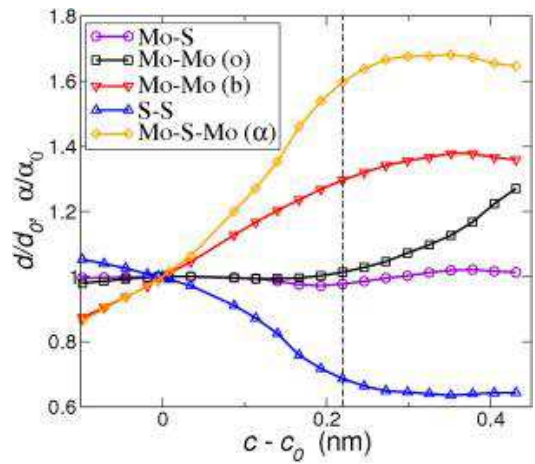}
\caption{(Color online) Variation of interatomic distances $d$ and Mo-S-Mo
angle $\protect\alpha $ upon stretching of $\mathrm{Mo_6S_3I_6}$
nanowires. Vertical dashed line shows the position of the second energy
minimum. Mo-Mo (o) and (b) denote the Mo-Mo distances in the octahedron and
across the bridge, respectively. The weakest segments where the wires break
are the Mo octahedra.}
\label{stretch167}
\end{figure}

We investigated the deformation of the $\mathrm{Mo_6S_3I_6}$ chains upon
longitudinal tensile stress by keeping the lateral lattice constant $a$
constant, to see how do the chains stretch and where are the weakest points
in the chains. The corresponding calculated stress-strain diagram is shown
in Fig.~\ref{stress167}(b). Initially, most of the strain is connected to
the deformation of the Mo-S-Mo bond angles across the bridging plane like in
an accordion while the Mo-S distances and the structure of the Mo octahedra
remain almost constant, (see Fig.~\ref{stretch167}). In this region the
Young modulus is $c_{11}=82$ GPa. Very soon the value of the critical
stress, $\sigma _{c}=0.39$ GPa is reached, (see Fig.~\ref{stress167}(a)),
when it becomes energetically more favourable for the Mo-S-Mo bonds to jump
into their stretched configurations. However, the potential barrier between
the two configurations prevents such transitions, the system remains in a
metastable state (dashed line in Fig.~\ref{stress167}) until the unstable
region is reached when all the Mo-S-Mo bonds suddenly jump at constant
stress to the stretched state. On contraction, upon reducing the stress, the
Mo-S-Mo bonds remain in the stretched state down into the metastable region
until they eventually jump into the slightly compressed state in the
unstretched state. The $\mathrm{Mo_6S_3I_6 } $ nanowires thus show a
hysteresis in the stress-strain graph, caused by the potential barrier
between the two energy minima which originate in the nature of the Mo-S-Mo
bonds across the bridging planes.

Of course, the Mo-S-Mo angles cannot open beyond $180^{\circ }$, and this is
the reason for an increase in $\sigma _{11}$ at large strains and for a
higher elastic constant, $c_{11}=94$ GPa, than in the first energy minimum.
The non-linear regime at higher $\epsilon _{11}$ is the consequence of
anharmonic forces and is not related to plastic deformation which is absent
in our DFT calculations. Of course, the chains eventually break at the
maximal stress. This defines the ideal tensile strength, $\sigma _{\max
}\approx 8.4$ GPa. Inspection of Fig.~\ref{stretch167} tells us that the
weakest links, where the chains break, are -- surprisingly -- the Mo
octahedra and not the Mo-S-Mo bridges. This is also the reason why all MoSI$%
_x$ nanowires have similar tensile strengths.

In the experiments the wires were subject to an initial stress which most
probably brings all the Mo-S-Mo bridges into their stretched configurations.
Only after this initial treatment the elastic constants were measured.
Therefore one must compare the experimental elastic constant with that of
stretched Mo-S-Mo bridge configurations. However, it remains a mystery, why
the experimental longitudinal Young moduli exceed the elastic constants 
$c_{11}$ reported in this paper by a factor of 4 or more.\cite{Kis}
The above values of $c_{11}$ were calcualted on a hexagonal array of molecular
chanin with all chains carrying equal load. For comparison, the Young modulus
of a hexagonal array of single-wall carbon nanotubes (SWCNT) with the lattice 
spacing 1.7 nm is $\sim 600$ GPa which is about 6 times the
calculated value of $c_{11}$.\cite{YFAR}

The accordion effect is present also in other \textit{one-dimensional} carbon 
chains like polyethylene or polyacetylene. What is unique to $\mathrm{%
Mo_{6}S_{4}I_{6}}$ is the double energy minimum. 
An additional sulfur in the centre of the bridging planes in $\mathrm{%
Mo_{6}S_{4}I_{6}}$ significantly increases the elastic constant ($c_{11}=114$
GPa), because it hinders the \textquotedblleft accordion
effect\textquotedblright\ of the Mo-S-Mo bonds. In case of $\mathrm{%
Mo_{6}S_{4}I_{6}}$ there is also no sign of a second energy minimum in
strained nanowires. 
Finally we notice that the calculated elastic constant $c_{11}$ of $\mathrm{%
Mo_{6}S_{3}I_{6}}$ is more than 3 times smaller than $c_{11}$ of $\mathrm{%
Mo_{6}S_{6}}$ or $\mathrm{K_{2}Mo_{6}S_{6}}$ where the flexible Mo-S-Mo
bridges are absent.\cite{Vil06}


\subsection{Electronic properties}

\begin{figure}[tbp]
\includegraphics[width=0.95\columnwidth]{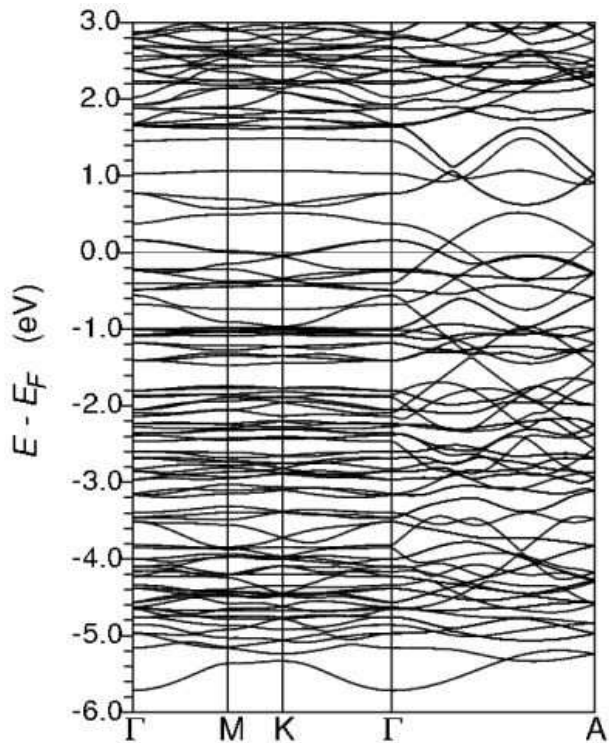}
\caption{Bandstructure of the conduction (valence) band along the special
lines of the hexagonal Brillouin zone. Three sub-bands (one of them doubly
degenerate) cross the Fermi energy and contribute to the charge carrier
transport.}
\label{spagh167}
\end{figure}

\begin{figure}[tbp]
\includegraphics[width=1.0\columnwidth]{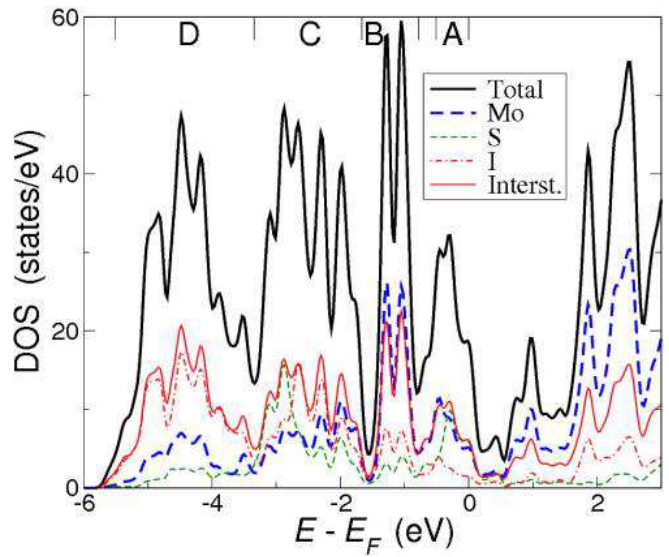}
\caption{(Color online) Density of states in the conduction (valence) band.
For Mo, S and I the partial charges in the muffin-tin spheres are plotted,
the rest are in the interstitial region. The main contributions to the
conduction band are from Mo-$4d$, S-$3p$ and I-$5p$ orbitals. The letters at
the top denote the energy ranges, discussed in the text.}
\label{dos167}
\end{figure}

\begin{figure}[tbp]
\includegraphics[width=0.6\columnwidth]{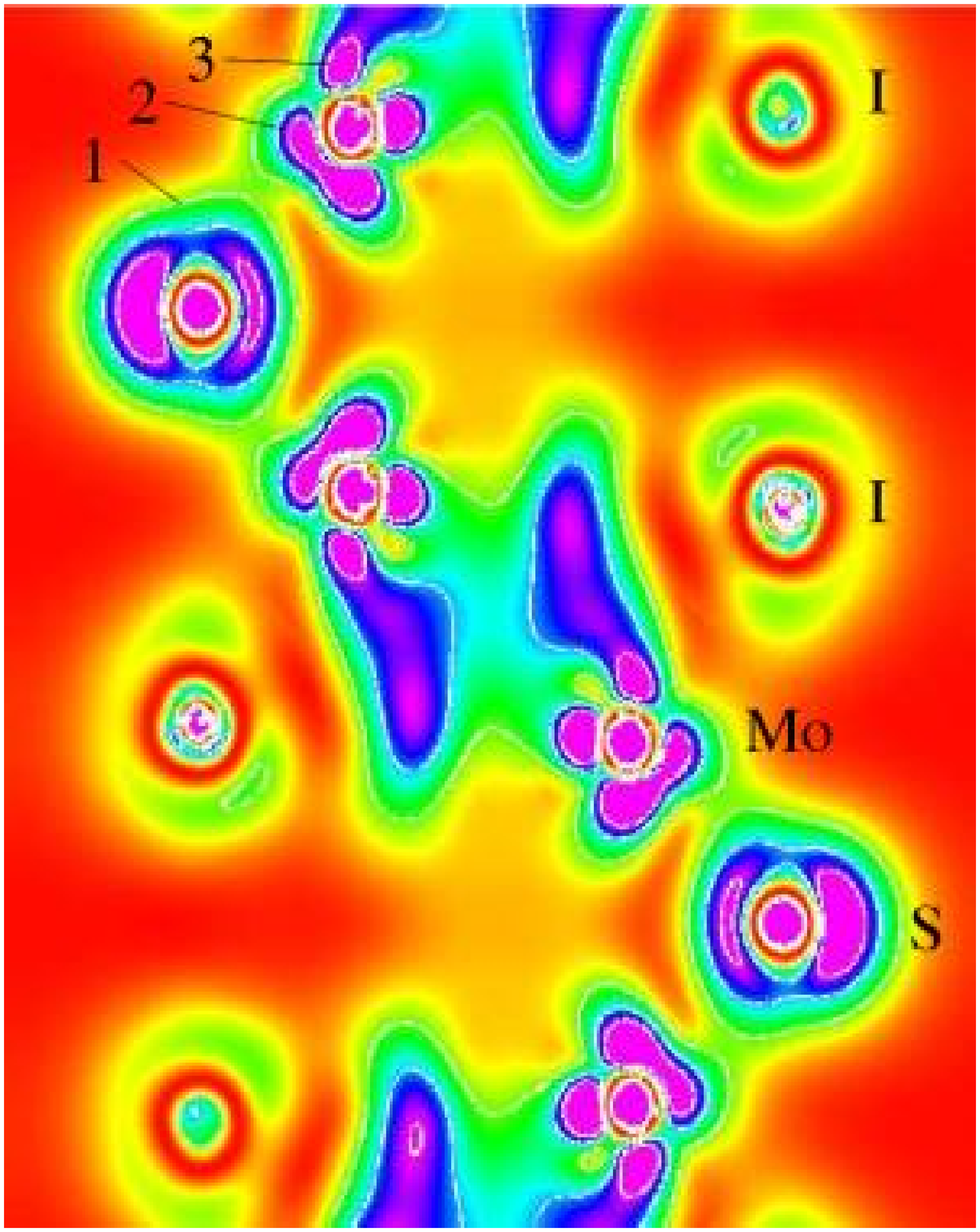}\newline
(a) \vskip5mm \includegraphics[width=0.8\columnwidth]{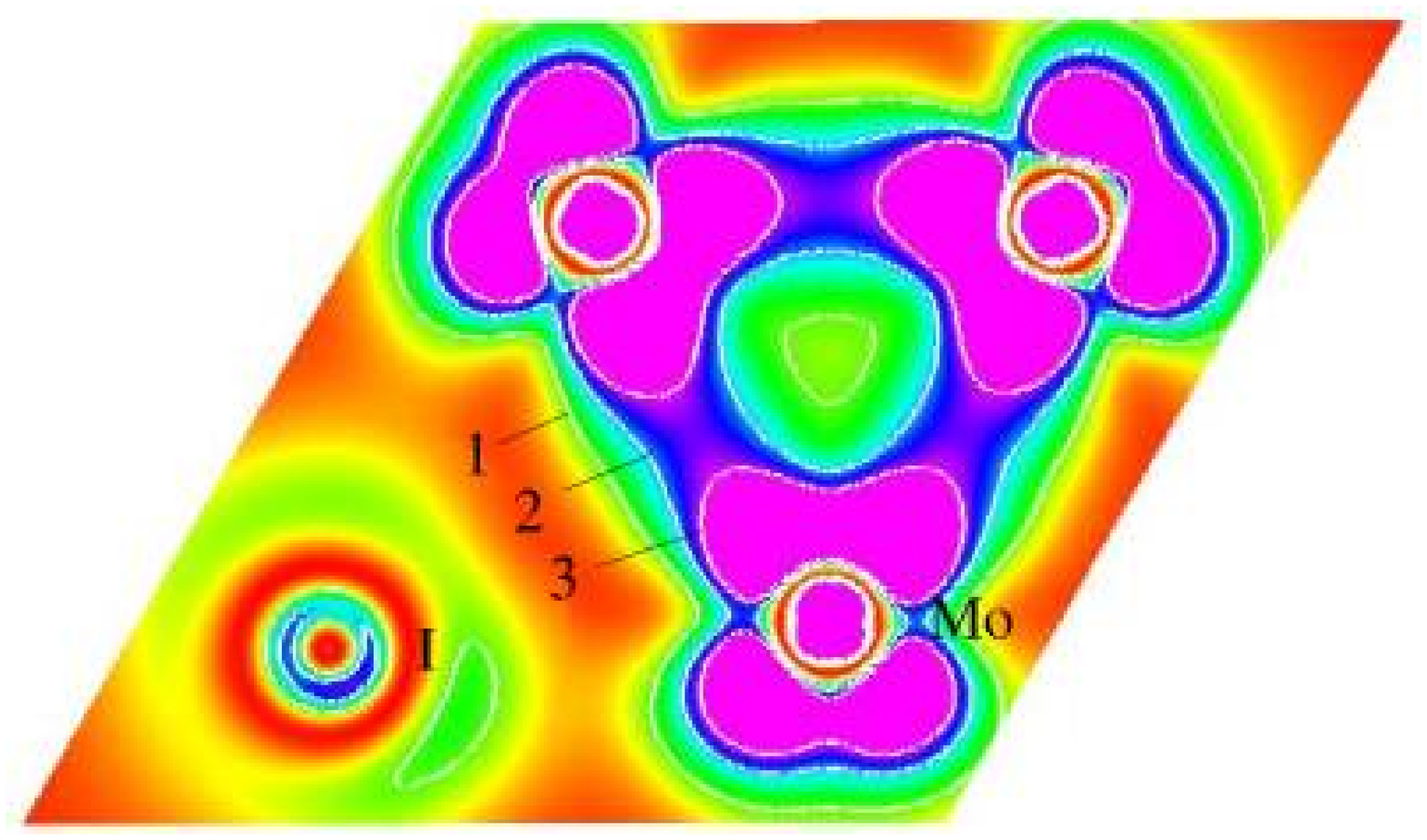}%
\newline
(b)\vskip5mm \includegraphics[width=0.5\columnwidth]{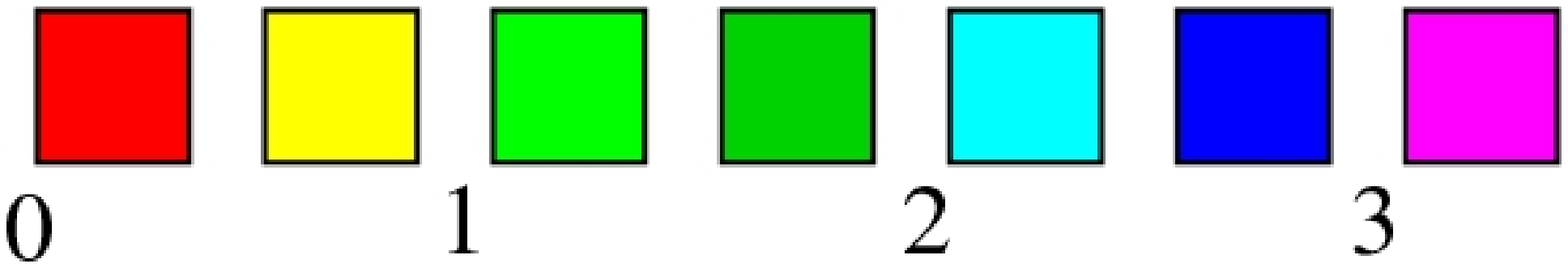}
\caption{(Color online) Valence charge density in the energy window A on the
(1100) plane through the centre of a $\mathrm{Mo_6S_3I_6 } $ nanowire (a)
and on the (0001) plane through Mo atoms (b). The isolines and the color
coding are in units of 0.0167 electrons/{\AA }$^3$. Clearly seen are the
charge concentration on Mo octahedra and through the bridging sulfur atoms.}
\label{dens167A}
\end{figure}

The electron bandstructure of the conduction or valence band, Fig. \ref%
{spagh167}, shows several interpenetrating sub-bands, three of them crossing
the Fermi energy. Because of the anisotropy in the crystal structure the
dispersion in the lateral directions is very small, we can classify 
Mo$_{6}$S$_{3}$I$_{6}$ as a quasi-one-dimensional conductor. 
The conduction band consists of hybridized
Mo-$4d$, S-$3p$ and I-$5p$ orbitals. The character of the sub-bands is most
clearly seen in the electron density of states (DOS), Fig.~\ref{dos167}.
Close to $E_{F}$ (region A in Fig.~\ref{dos167}), the hybridization of the
Mo-$4d$ and S-$3p$ states is the origin of good conduction along the Mo-S-I
chains. This is not surprising since a conduction electron on its way along
the chain must pass the sulfur bridges (S-$3p$ states). This is also seen in
the real-space charge-density plot in Fig.~\ref{dens167A}(a) where the
charge density in the energy interval A 
is seen to run continuously throughout the chain. A cross-cut through the Mo
(0001) plane, Fig.~\ref{dens167A}(b), reveals a combination of metallic
bonds (with a uniform distribution of electrons within the Mo octahedron)
and covalent bonds between neighbouring Mo atoms (a covalent bond is
directional between the atoms). The I atoms, which are close to this plane,
are only very weakly bound to the Mo octahedron in this energy window. Lower
in energy, i.e., in energy window B, a strong Mo-$4d$ double peak, also
responsible for the Mo-Mo bonding, dominates the DOS.


Towards the bottom of the conduction band (energy interval D) the DOS is
dominated by the states belonging to the dressing atoms (I-$p$ in our case
and S-$p$ in case of $\mathrm{Mo_6S_6}$\cite{Vil06}) and to polar covalent
bonding between Mo and I.

The regions C and D of the conduction-valence band are mainly responsible
for the Mo-S and Mo-I bonding. We identify these two bonds as polar covalent
because of partial ionization of I and S. In fact, the charges of the
muffin-tin spheres are as follows: Each Mo gives away 3.5 electrons, S 0.9
electrons and I 2 electrons, all these electrons fill the interstitial
region. For comparison, Mo in metallic state gives away (into the conduction
band) only two electrons. 
These charges are also in accordance with electronegativity. For a Mo-S bond
the difference in electronegativity is 0.42 on the Pauling scale whereas for
a Mo-I bond it is 0.50, I is slightly more negatively charged than S. 

Comparison with the DOS of $\mathrm{Mo_6S_6}$ chains which have no S bridges,%
\cite{Vil06} shows similar occupations of Mo-$4d$ states in both cases but a
much weaker occupation of the S-$3p$ states in case of $\mathrm{Mo_6S_6}$
close to $E_F$.

The bands below the conduction band are localized to the respective atoms
and have well-defined characters with very little hybridization. 
In the whole DOS we see very little s-p hybridization in the sulfur
orbitals from which we conclude that the s-p hybridization cannot be
responsible for the stabilization of the second energy minimum at $c=13.7$ {%
\AA } in $\mathrm{Mo_6S_3I_6}$.

From Fig.~\ref{spagh167} the estimated Fermi velocities are in the range
between 0.7 and 1.3 $\times 10^{6}$ m/s. With a scattering lifetime $\tau
\sim 0.7\times 10^{-14}$ s ($\Gamma =0.1$ eV) this leads to a mean free path
along the wires of the order 5 to 9 {nm}, enough to justify a delocalized
description of the charge carriers.

Comparison with the bandstructure of the $\mathrm{Mo_6S_3I_6}$ isomer with 
iodines in the bridging planes (spacegroup R3)\cite{MKPG05} shows that the
bandstructure is very sensitive to the anion occupancies, in particular to the 
occupancy of the bridging atomic positions. 
Although the present R$\overline{3}$c structure has virtually no calculated 
energy gap, the spacegroup R3 isomer has a $\sim 0.6$ eV bandgap above $E_F$, 
According to our DFT calculations both structures have very similar formation
energies, therefore we expect that both are synthesized simultaneously in comparable 
quantities. For the same reason we cannot exclude the presence of other 
$\mathrm{Mo_6S_3I_6}$ nanochains with more disorder in the anion occupancies.
The bandstructure, the corresponding optical spectra and the longitudinal
conductivity can be an effective criterion to distinguish between different isomers.

To gain insight on the conductivity behaviour we calculate the dielectric
tensor $\epsilon (\omega )$ in the random-phase approximation and in the
limit $q\rightarrow 0$.\cite{ADS06} The static metallic conductivity is
associated with the Drude peak which comes from transitions between states
close to $E_{F}$. From the Drude peak damped with $\Gamma =0.1$ eV we find
the longitudinal conductivity $\sigma _{||}\sim 5\times 10^{3}$ S/cm. This
is a factor $\sim 4$ lower than $\sigma _{||}$ of $\mathrm{Mo_{6}S_{6}}$
nanowires, which we attribute to the weaker conduction of S bridges. The
lateral conductivity is one order of magnitude smaller as well. This leads
to a very short carrier mean free path in the lateral directions. Clearly
the band picture breaks down in the lateral directions and the
charge-carrier transport is via hopping between neighbouring chains. Of
course, in real systems with finite-length chains the charge carriers must
hop also between neighbouring chains and this can also hinder the charge
carrier transport along the chains. 
In the calculation,  $\Gamma =0.1$ eV was taken arbitrarily -- for very 
clean nanowires, $\Gamma$ could be substantially less and the conductivity 
correspondingly higher. With $\Gamma =0.1$ the longitudinal conductivity 
of  $\mathrm{Mo_{6}S_{3}I_6}$  is about two order of magnitudes smaller than 
$\sigma$ reported for SWCNT at 300 K which are considered to be one of the
best one-dimensional conductors.\cite{TLN}


\subsection{Electronic properties of stretched nanowires}

\begin{figure}[tbp]
\begin{minipage} {4.25cm}
      \includegraphics[width=1.0\columnwidth]{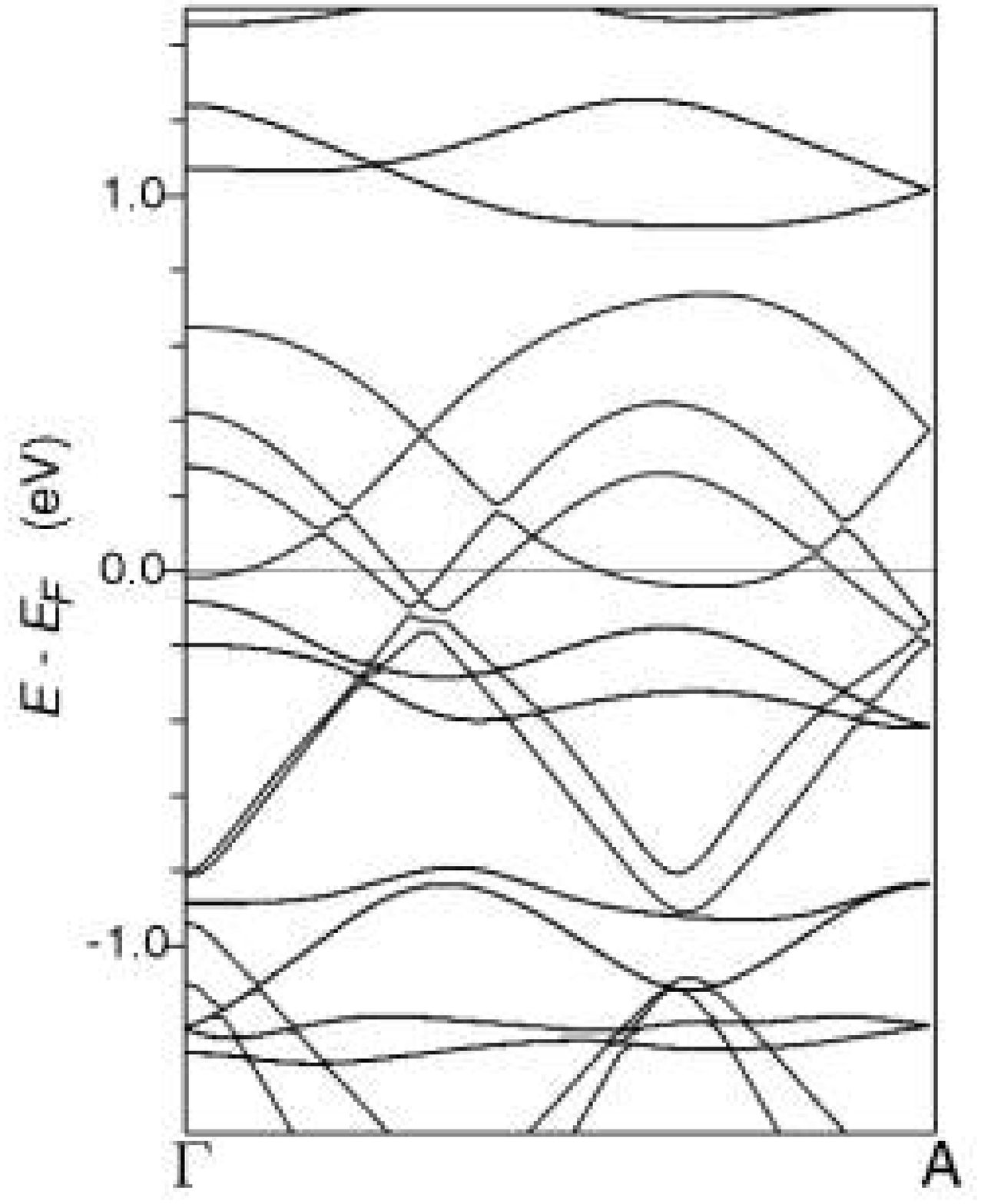}\vglue-2mm
             $c - c_0 = 0.114$ nm
   \end{minipage}
\begin{minipage} {4.25cm}
       \includegraphics[width=1.0\columnwidth]{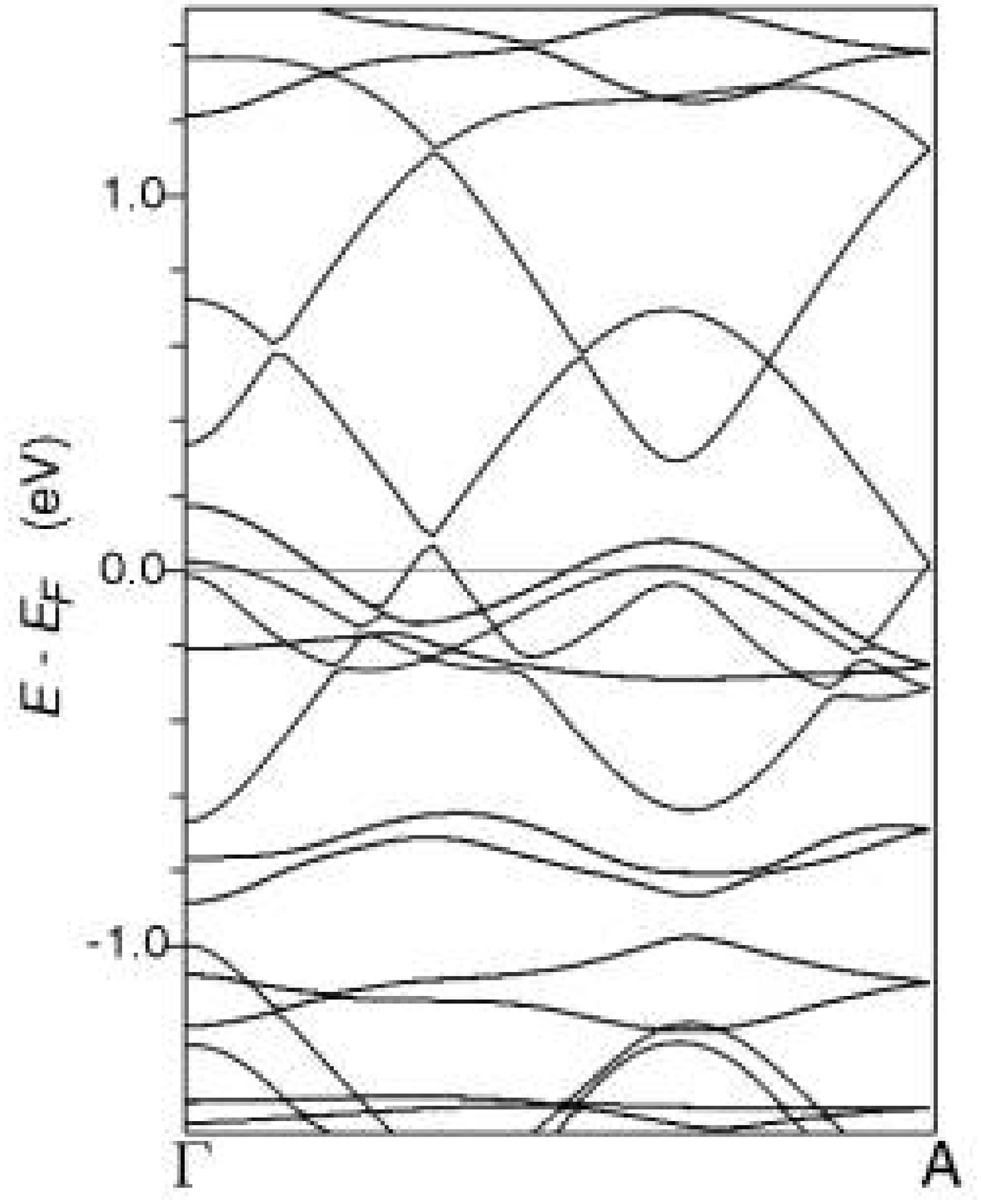}\vglue-2mm
    $c - c_0 = 0.167$ nm 
   \end{minipage}\newline
\begin{minipage} {4.25cm}
      \includegraphics[width=1.0\columnwidth]{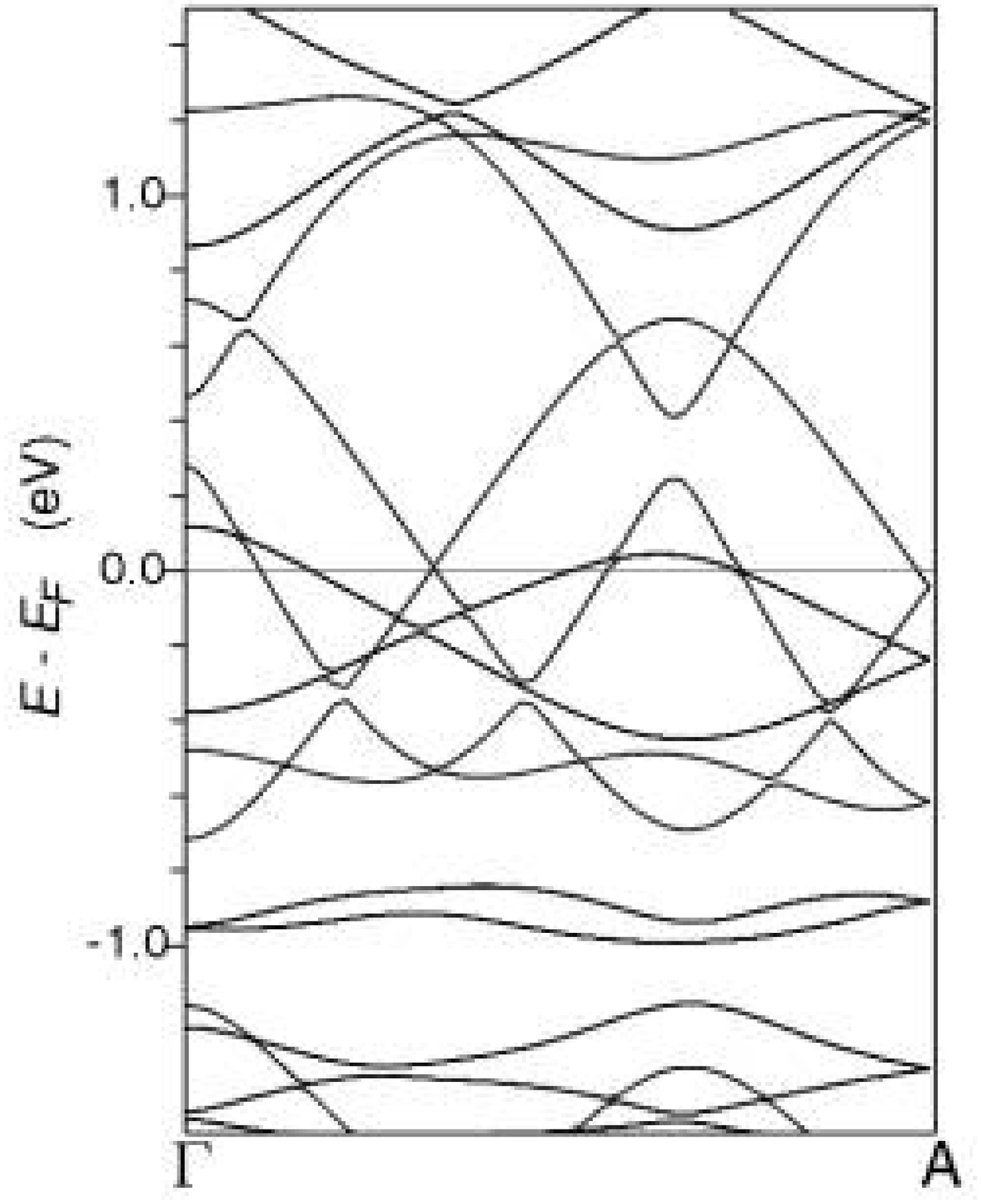}\vglue-2mm
      $c - c_0 =  0.220$ nm 
   \end{minipage}
\begin{minipage} {4.25cm}
      \includegraphics[width=1.0\columnwidth]{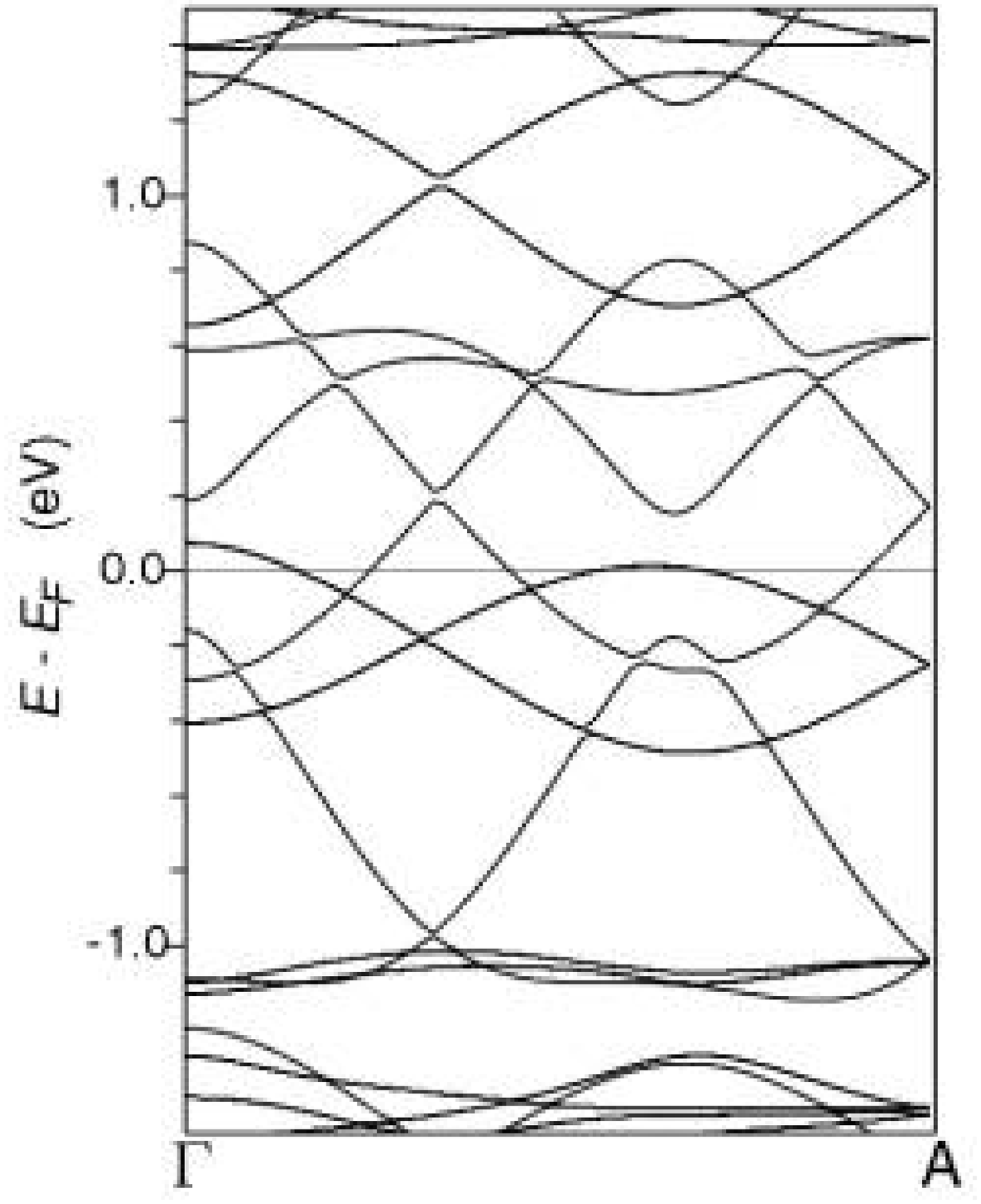}\vglue-2mm
     $c - c_0 = 0.325$ nm 
   \end{minipage}
\caption{Bandstructures of $\mathrm{Mo_6S_3I_6 } $ nanowires subjected to
longitudinal strain. Several sub-bands cross the Fermi energy and the width
of these sub-bands is maximal close to the second energy minimum at $c - c_0
= 0.220$ nm. }
\label{band_strain}
\end{figure}

\begin{figure}[tbp]
\includegraphics[width=0.45\columnwidth]{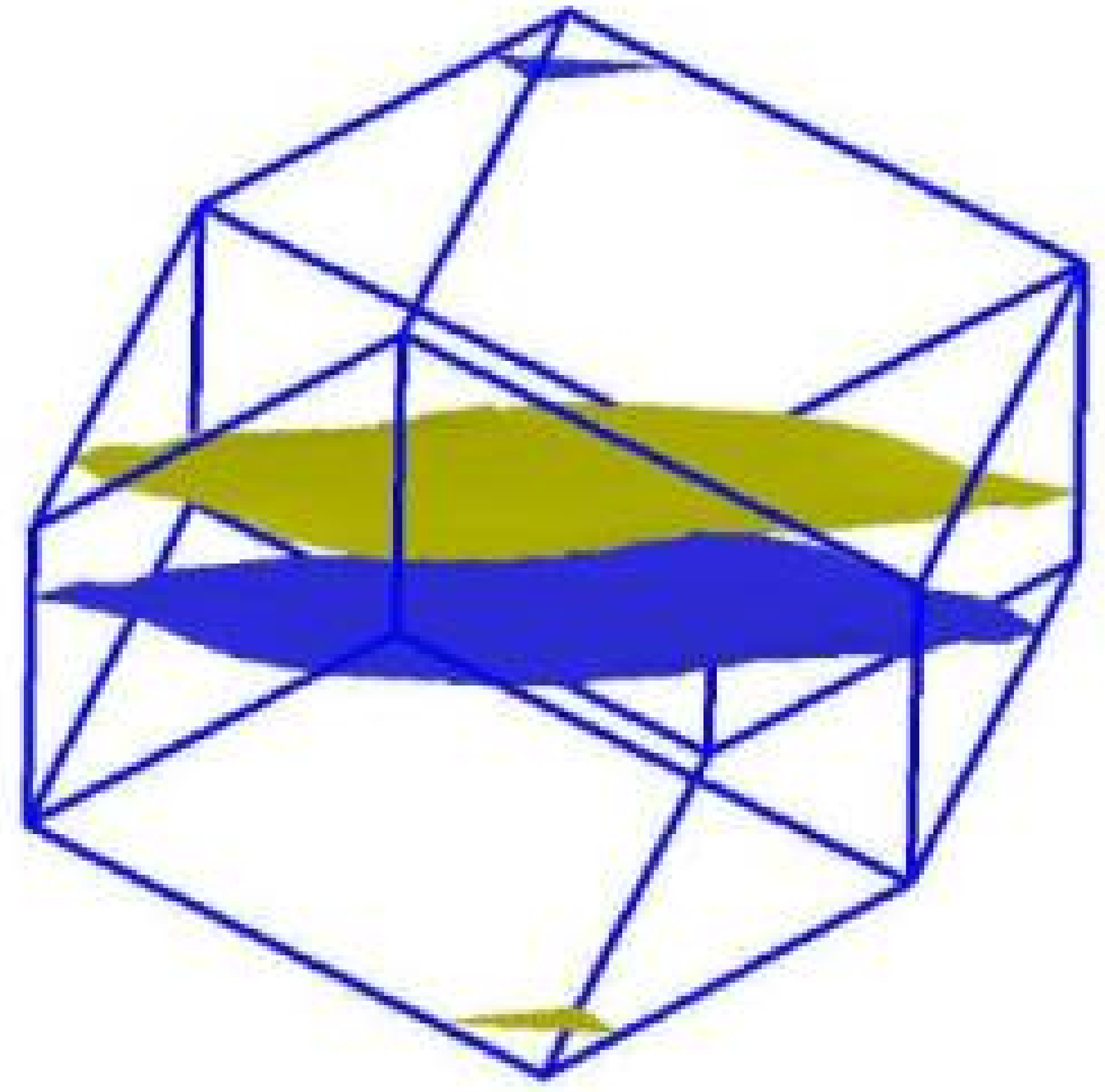} %
\includegraphics[width=0.45\columnwidth]{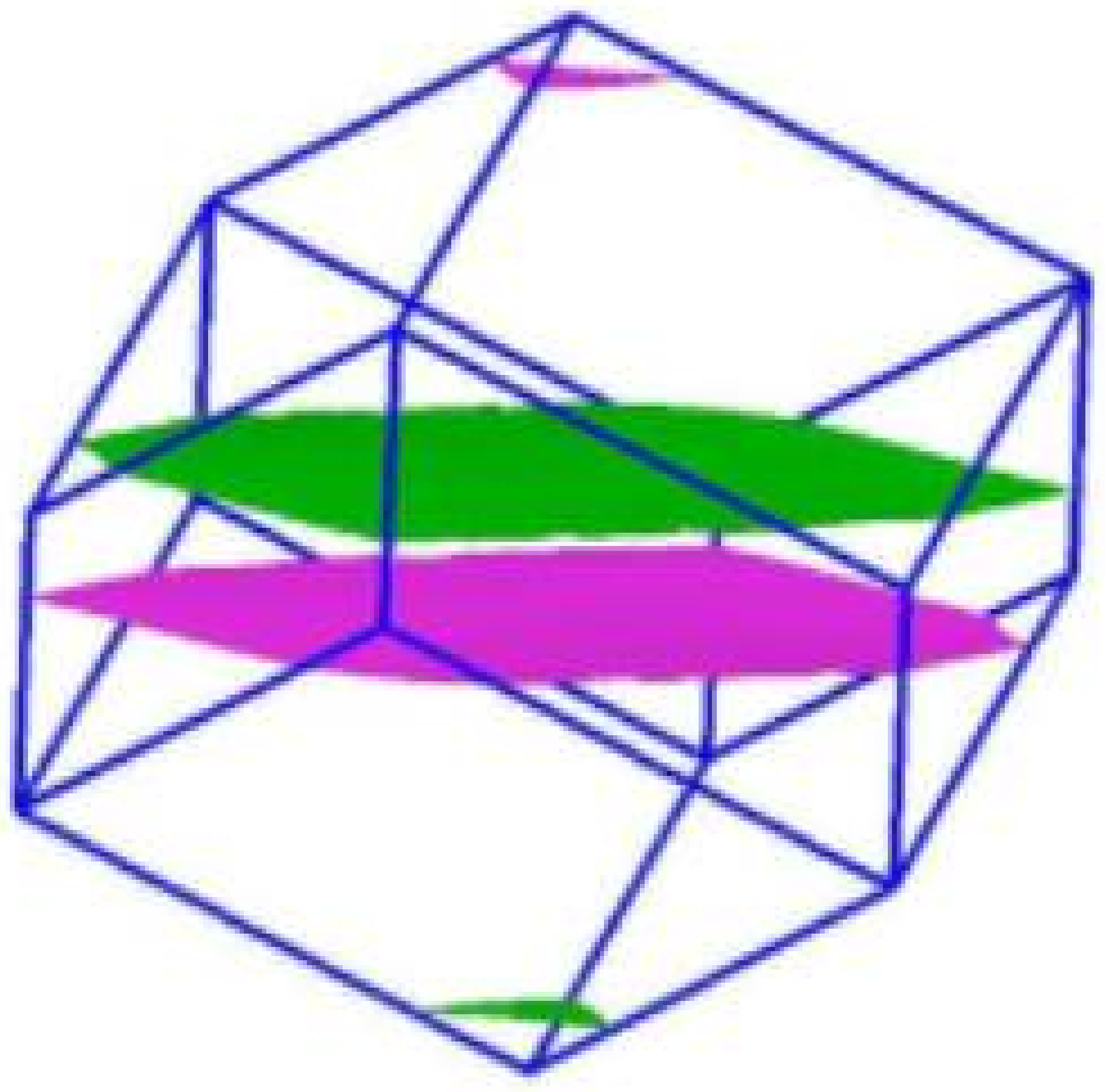}\newline
\includegraphics[width=0.45\columnwidth]{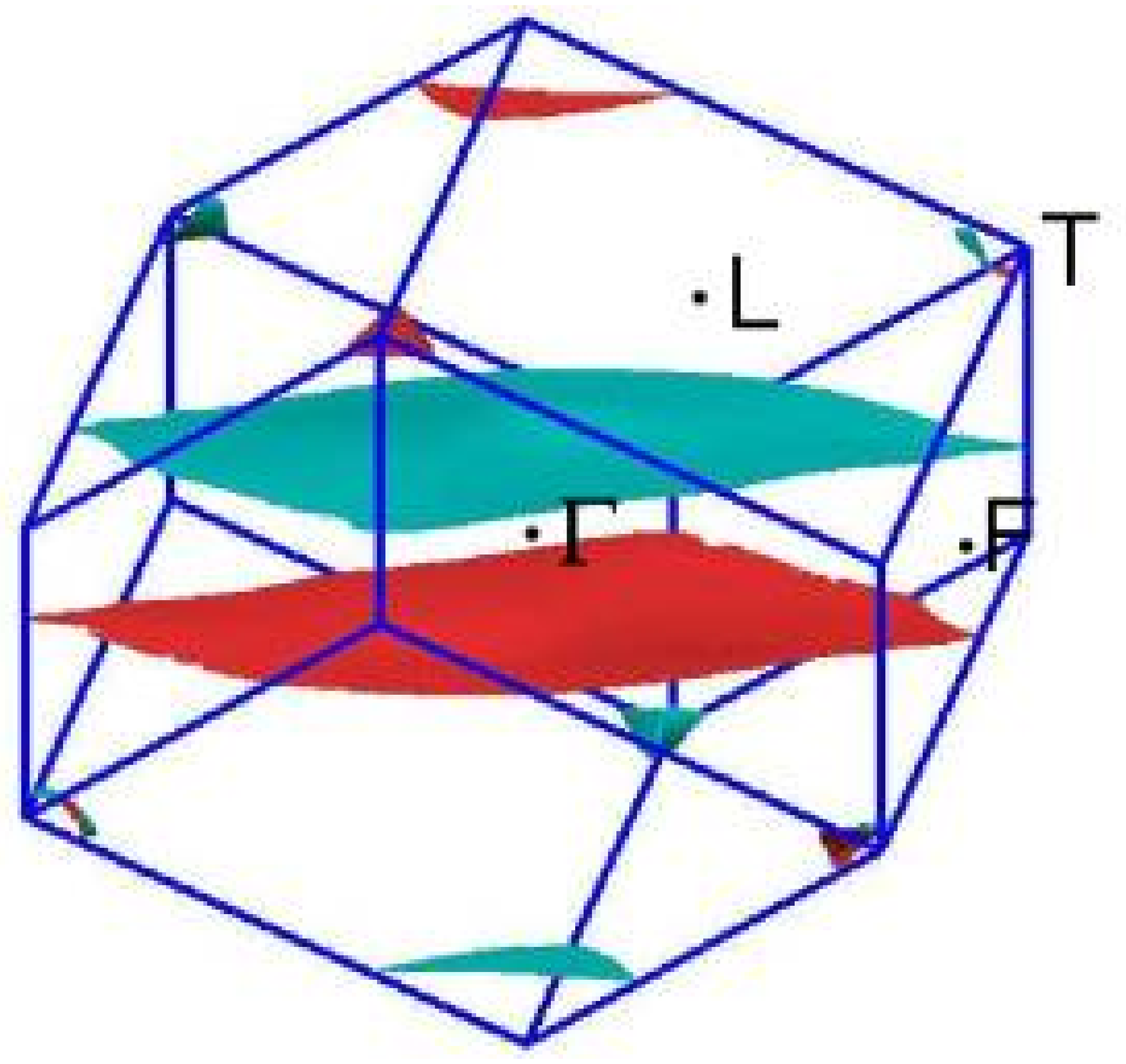} %
\includegraphics[width=0.45\columnwidth]{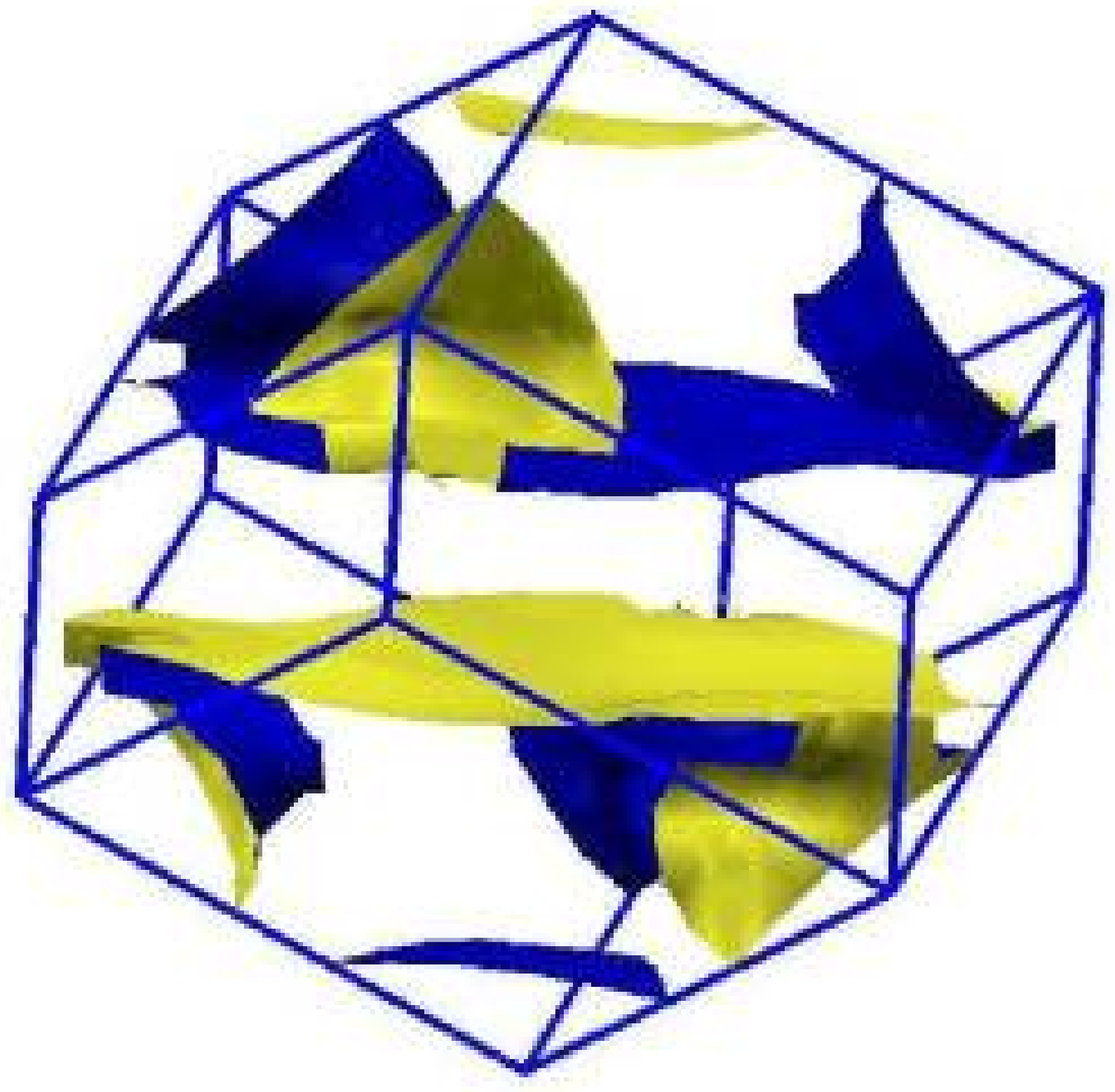}
\caption{Fermi surfaces and the Brillouin zone of the rhombohedral unit cell
of stretched nanowires ($c=13.76$ {\AA }). }
\label{fermi}
\end{figure}

The bandstructure of MoSI$_x$ nanowires and consequently also the Fermi
surfaces and electrical conductivity are extremely sensitive to strain along
the hexagonal $c$ axis. In Fig.~\ref{band_strain} we show the bandstructure
along the $[001]$ symmetry direction for four different strains. Although it
is difficult to follow a particular sub-band through the whole range of
strains, we observe that in general the bandwidth of the sub-bands that cross 
$E_{F}$ and which are responsible for the charge-carrier transport first
increases with strain up to the second energy minimum ($c=13.7$ {\AA }) and
then decreases. Not surprisingly, the bandwidth is related to the Mo-S
distance, Fig.~\ref{stretch167}, which reaches its minimum around $c=13.7$ {%
\AA }. Beyond $c=13.7$ {\AA } the Mo-S distance increases and the sub-band
width decreases, as expected. Similar behaviour is expected for the
longitudinal DC conductivity. Above 15.2 {\AA } 
$\sigma$ drops rapidly because the chains start to break apart. Assuming the
same $\Gamma$ in the lateral direction we find that the anisotropy in the
static conductivity is large, typically of the order 100 for stretched
nanowires. The Fermi surfaces of $\mathrm{Mo_6S_3I_6 } $ in the second
energy minimum ($c=13.76$ {\AA }, Fig.~\ref{fermi})\ show typical features
for a quasi-one-dimensional material, almost planar Fermi surfaces with very
little curvature. The number of Fermi surfaces and the Fermi velocities are
very sensitive to longitudinal strain, and so is the conductivity. 

The $\mathrm{Mo_{6}S_{3}I_{6}}$ nanowires thus behave as
quasi-one-dimensional metals in the whole range of strains. This is
different to the $\mathrm{Mo_{12}S_{9}I_{9}}$ nanowires where a transition
from metallic to narrow-gap semiconductor has been reported.\cite{YOBT06}

The lateral conductivity is dominated by hopping to neighbournig chains
through the bridging sulfur atoms, see Fig.\ref{dens167A}. Upon longitudinal
strain, the sulfur atoms move towards the centre of the wires whereas the
interchain separation, dominated by the non-conducting I atoms, remains
almost constant. As a consequence, the lateral conductivity drops rapidly
with tensile strain.

The variation of the conductivity with strain is described with the strain
gauge factor, $G=\Delta \rho /(\rho \Delta \epsilon )$ where $\rho =1/\sigma 
$. For the longitudinal conductivity we find $G$ of the order -10 which is
somewhere in between the typical values for semiconductors and for metals.
However, the unusual feature of MoSI$_x$ nanowires, and great potential
advantage in applications is their extreme flexibility (large $\Delta
\epsilon $) arising from the accordion effect in the Mo-S-Mo bridges. Values
of $\Delta \sigma /\sigma \sim 1$ can be reached in the elastic regime which
is an order of magnitude better than $\Delta \sigma /\sigma$ in classical
metallic strain gauges. 


\section{Discussion and Conclusion}

The $\mathrm{Mo_{6}S_{3}I_{6}}$ nanowires investigated in this paper have
some similarities to the skeletally equivalent $\mathrm{Mo_{12}S_{9}I_{9}}$,
but also some important differences. Both nanowires have three sulfur atoms
in the bridging planes between the Mo octahedra and each bridging plane
shows two energy minima upon uniaxial strain, one in the ground state
(unstretched wire) and one corresponding to the stretched wire. 
The potential barrier between the two minima is high and
prevents thermally excited transitions between the two minima, it is
responsible for hysteresis in the stress-strain behaviour of $\mathrm{%
Mo_{6}S_{3}I_{6}}$. This interpretation is slightly different from Yang et
al.\cite{YOBT06} where the initial atomic configurations were subjected to
random distortions, which of course mimic the entropy which helps to
overcome the potential barriers in the metastable region.

Although the Mo-S-Mo bonds in $\mathrm{Mo_{6}S_{3}I_{6}}$ nanowires are
very deformable, the Young modulus of the unstrained energy minimum is only
slightly smaller than Y of the second minimum with strained Mo-S-Mo bonds.
The lack of these very deformable Mo-S-Mo bridges in $\mathrm{Mo_{6}S_{6}}$
is also the reason why the latter chains have a higher Young modulus.
Surprisingly, the chains break in the metallic-covalent bonds between Mo in
octahedra and not, as one would expect, in the polar covalent Mo-S bonds
across the S bridges. This is the reason why all Mo nanowires, dressed with
I or S anions and with or without bridging atoms, have similar calculated
tensile strengths. However, the much larger experimental Young's moduli
reported by Kis et al.\cite{Kis} remain a puzzle at present. The $\mathrm{%
Mo_{6}S_{3}I_{6}}$ nanowires behave as quasi- one-dimensional conductors in
the whole range of investigated strains. The conductivity involves the Mo
octahedra and the bridging atoms and is extremely sensitive to strain,
making this material very suitable for stain gauges.

The fact that the experimentally reported shear modulus appears to be an
order of magnitude smaller than in SWCNTs suggests that 1D effects will be
even more pronounced than in SWCNT ropes, and the low-energy spectrum as $%
E\rightarrow E_{F}$ is expected to be modified as a result. Very clean MoSI$%
_{x}$ nanowires with good contacts may be expected to behave as ballistic
quantum wires over lengths of several $\mu $m. On the other hand, with
high-impedance contacts MoSI$_{x}$ nanowires may be thought of excellent
candidates for the observation of Luttinger liquid behaviour\cite%
{MatveevGlazman,Kim}. Of course, one-dimensional metallic systems such as Mo$%
_{6}$S$_{3}$I$_{6}$ are subject to strong localisation effects. Defects on a
molecular wire can effectively stop and localise electrons, whereupon
hopping between molecular chains becomes relevant. The pronounced 1D nature
of the nanowires makes them a uniquely versatile and user-friendly system
for the investigation of 1D physics.


\begin{acknowledgments}
We would like to acknowledge interesting discussions with A. Meden.
This work was supported by the Slovenian Research Agency under the contract
P1-0044. The crystal structures were visualized by Xcrysden.\cite{K99}
\end{acknowledgments}


\end{document}